# π-Stacking in Charged Thiophene Oligomers

*Damián A. Scherlis\* and Nicola Marzari*

Department of Materials Science and Engineering and Institute for Soldier Nanotechnologies,

Massachusetts Institute of Technology, Cambridge MA 02139

damians@mit.edu, marzari@mit.edu

**ABSTRACT.** The π-stacking of oxidized thiophene oligomers is investigated using extensive ab-initio quantum chemistry methods. Dimers of singly-charged oligothiophenes are found to be unstable in the gas phase, but can be stabilized as bound dications in the singlet state by a polarizable solvent such as acetonitrile. Our calculations provide a detailed description of the mechanisms and the energetics involved in the dimerization phenomenon, and highlight the role and importance of the environment in the stabilization of the stacks. The need for accurate treatments of electronic correlations and of solvation effects for a realistic description of these materials is underscored.

## Introduction

The promising electronic properties of organic conducting materials, and in particular of thiophene oligomers and polymers, have prompted a vast amount of basic research[1] and a growing number of developments in the field of molecular electronics.[2] Applications run from light-emitting diodes[2a] to thin film transistors[2b] and sensing and actuating materials.[2c] The characterization of the electronic structure for the ground and excited states, of the thermodynamical stability, and of charge transport mechanisms such as interchain hopping[1c,f] are essential to engineer thiophene-based devices. Experimental evidence has long accumulated suggesting reversible dimerization in assemblies of oxidized oligothiophenes.[1a-e] The formation of such aggregates of two or more cations due to $\pi$-$\pi$ interactions—so called $\pi$-stacking—is fundamental in determining equilibrium geometries and conductivities in solution or in the solid state.[1c]

In the dimerization of cation radicals, intermolecular attraction forces must overcome Coulombic repulsion, and the net outcome is the result of a delicate balance also dependent on the length of the oligomer and the polarity of the surrounding medium.[1b,c] The pairing of charged (anion) radicals was first reported for diphenylethylene[3a] and for tetracyanoquinodimethane[3b] (TCNQ) four decades ago, and the dimerization of cationic and anionic species in solution and in the solid state has been extensively studied ever since.[3c-f] Although early theoretical work successfully rationalized this phenomenon using model Hamiltonians and semiempirical approaches,[3g-i] the study of the structural and thermodynamic aspects of the process has continued till the present day.[4] The identification of intrinsic factors— electrical charge, electron delocalization, $\pi$-$\pi$ bonding—and those of extrinsic origin—packing interactions, counterions, solvent effects—is still a fundamental challenge. In particular, for the case of oligothiophenes and in spite of their prominence gained in materials applications, a thorough knowledge on the dimerization process is missing. Quantum calculations in vacuum and in solution provide a unique tool to identify and measure these effects.

In the dimerization of neutral monothiophenes, dispersive forces give rise to long-ranged interactions between the aromatic rings, leading to modest binding energies[5] (~2-3 kcal/mol). In the case of oxidized aggregates, however, the picture is far from conclusive. In an elementary molecular-orbital model, semi-occupied HOMOs combine upon dimerization to form occupied bonding and empty antibonding orbitals, counterbalancing the electrostatic repulsion. Stability is ultimately determined by the subtle differences between these two opposing contributions. A density-functional theory (DFT) study of singly-charged quaterthiophene radicals[6] showed strikingly different outcomes depending on the exchange-correlation functional chosen. A deep local minimum is found in LDA in the energy of the doubly-charged quaterthiophene dimer as a function of separation; on the contrary, GGA-PW91 does not support the existence of stable or metastable bound states. Even though overbinding in LDA is often observed, such a dramatic disagreement is unusual, and questions the nature or existence of dimer stacking. We present here a comprehensive study of the interactions between singly-charged thiophene oligomers in vacuum and in solution, to assess the existence of stable or metastable dimerized bound states and the role played by the environment.

**Computational Approach**

**General Remarks.** Calculations were performed with the Gaussian 98 program.[7] The basis sets correspond to those implemented in the software, unless otherwise noted. The stability of the wavefunctions was checked for every energy minimization. The basis set superposition error (BSSE) was computed through the counterpoise method[8] to correct the interaction energies in all cases, with the exception of the CASSCF and the IPCM calculations (see below). The geometries of the isolated bi-, ter-, and quaterthiophene cations were optimized with MP2 at the 6-31G(d) level, and the relaxed structures used for the calculations of the dimers. The optimal parallel alignment between monomers was determined by constrained minimizations with the planes of the molecules fixed at a distance of 3.4 Å. Both singlet and triplet states were computed, and the lowest state was picked to construct the interaction energy curves. In general, regardless of the method, we found that up to 4.0 Å of separation the ground state is a singlet, while for longer distances it becomes a triplet.

**Hartree-Fock and MP2 Calculations.** Unrestricted Hartree-Fock calculations have been performed for the bithiophene dimer employing the 6-311+G(d,p) basis set. For distances shorter than 4.0 Å, the closed-shell wavefunction is found to be unstable. The quadratic convergence procedure leads to a stable ground state exhibiting open-shell singlet character: the unpaired spin density ($\int \rho_\alpha - \rho_\beta \, d\mathbf{r}$) integrates to +1/-1 electron on either monomer, whereas the $S^2$ value is close to 2. For interchain distances beyond 4.0 Å the triplet is the ground state. For the second-order Møller-Plesset perturbation (MP2) method we used three different basis sets: 6-311+G(d,p), 6-311G(d), and 6-31G*(0.25).[9]

**CASSCF Calculations.** Complete active-space multiconfiguration SCF calculations (CASSCF) were performed on the bithiophene dimer. An active space of six electrons and four orbitals along with a 6-311+G(d) basis were employed. Comparisons between CASSCF(10,8), CASSCF(8,8) and CASSCF(6,4) calculations with the smaller basis 6-31G(d,p) show that the enlargement of the active space provides only marginal improvements to the interaction energy, at the expense of a dramatic raise in the computational costs. On the other hand, even if a small active space may not be enough to recover the total correlation, we expect it will account for most of the relative change in the correlation energy associated to the formation of the dimer. Therefore, we chose to keep a configuration space of moderate size while enlarging the basis set, on which the correlation energy exhibits a stronger dependence. In order to implement basis-set superposition corrections, CASSCF calculations should be performed on the charged bithiophene monomer, which has as many basis functions as the dimer, but half the number of electrons. On the other hand, while the dimer is a closed shell system, the monomer is a radical, and this makes the application of the counterpoise method unfeasible due to the strongly non-equivalent configuration spaces. To compute the interaction curve, the energy corresponding to two charged bithiophenes laying 100 Å apart (minus the electrostatic contribution) was subtracted from the energies of the π-dimers. This produces a balanced energy difference, where the number of configurations in products and reactants is roughly the same (not exactly the same, though, because the multiplicity of each of these systems is different).

**Density Functional Calculations.** Calculations were performed within the local-density and the generalized-gradient approximations (LDA and GGA). The Vosko-Wilk-Nusair parametrization[10] was adopted for LDA, and the PW91 exchange-correlation functional for GGA.[11] Additional DFT and hybrid DFT/Hartree-Fock simulations for the bithiophene dimer were performed at the BLYP[12] and the B3LYP[13] levels, in order to assess the performance of these functionals. We also examined the modified PW91 functional proposed by Adamo and Barone[14] (hereafter denoted MPW91) specifically adapted to describe van der Waals interactions. In the case of the quaterthiophene dimer, we used the double-zeta valence plus polarization (DZVP) basis sets designed for DFT by Godbout and coworkers.[15]

**Density-Functional Calculations within the Isodensity Surface Polarizable Continuum Model (IPCM).** To model the effect of solvation, we performed DFT simulations at the PW91/6-311+G(d,p) level with the dimer immersed in a polarizable dielectric continuum, using the IPCM method.[16] In this approach the solute is enclosed in an isodensity surface, which was chosen to be 0.0004e in our calculations. Inside the solute cavity the dielectric constant is 1, while outside it is set to the value corresponding to bulk acetonitrile (36.64). The charge density of the solute polarizes the solvent interface, which in turn affects the electronic density of the dimer. The polarization of the solvent and the electronic density are optimized self-consistently. The effect of outlying charge is accounted for by means of an additional effective charge, distributed according to the solute electronic density. The basis superposition error was not computed. The addition of the ghosts atoms required in the counterpoise method introduces basis functions centered in the solvent region, and that interferes with the definition of the isodensity surface. However, we expect the BSSE to be as low as tenths of kcal/mol, since the corrections found in vacuum with PW91/6-311+G(d,p) remained well under 1 kcal/mol.

**Results and Discussion**

In order to study the nature and strength of the interaction between charged oligothiophenes, we first investigated the pairing of singly-charged quaterthiophenes facing each other at a separation of 3.4 Å, using GGA-PW91. Two minima are found when one unit is shifted along its plane in the direction of its molecular axis. One local minimum corresponds to an eclipsed geometry with the quaterthiophenes exactly opposing each other, while a slightly lower minimum is found after a displacement of 2.0 Å, when quaterthiophenes assume a "slipped" configuration, as depicted on top of Figure 1a (such arrangement has been already observed experimentally in the solid state[1d] for a terthiophene derivative, and for dimers of charged conjugated and aromatic compounds[4]). The slipped configuration has been used to obtain all curves shown in Figure 1a, where the binding energy is plotted as a function of the distance between molecular planes. We do reproduce the striking disagreement between LDA and GGA-PW91 (Figure 1a), in agreement with previous findings[6]: a deep metastable minimum is present in LDA (nearly 17 kcal/mol) and disappears in GGA-PW91. In order to assess the origin and stability of stacking in these aromatic compounds, and to resolve such remarkable discrepancy, we have applied higher-level correlated approaches (such as CASSCF and MP2) to the case of charged bithiophene dimers. Bithiophenes were chosen because their moderate size allows computationally demanding treatments, while exhibiting the same DFT discrepancy found in longer oligomers.

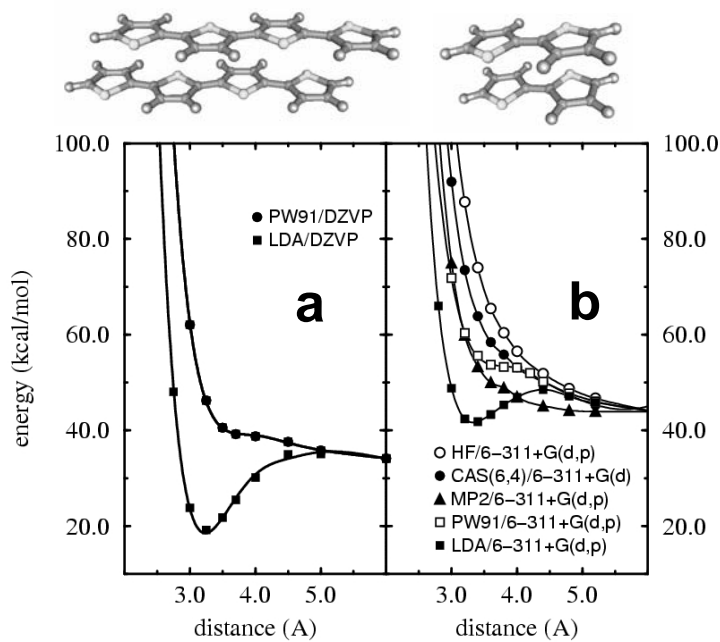

**Figure 1.** Interaction energy as a function of separation between two singly-charged oligothiophene cations in vacuum, obtained with different quantum chemistry methods. (a) quaterthiophene, (b) bithiophene.

A comparative summary of all methods is shown in Figure 1b, with binding curves for the eclipsed bithiophene dimers as obtained from Hartree-Fock, DFT, MP2, and CASSCF calculations. BLYP and B3LYP, shown separately in Figure 2 for more clarity, overlap with CASSCF. Also in Figure 2, results for the MPW91 functional (a version of PW91 modified for van der Waals complexes[14]) are included, showing a profile that parallels PW91, with slightly higher interaction energies.

In the past few years MP2 has been successfully applied to investigate the stacking of neutral aromatic molecules.[17,18] Studies on neutral systems have pointed out that MP2 has a tendency to overestimate the correlation interaction energy as the basis approaches saturation; this trend is somewhat balanced by the use of moderate size basis sets as the 6-31G*(0.25), which has shown to provide geometries and interaction energies in agreement with higher order methods as CCSD(T).[18,19] While in the neutral case the forces between chains are predominantly van der Waals, in the charged dimers the main contributions come from electrostatic and covalent terms. As a consequence, the long-range terms play a secondary role in the net interaction between charged oligomers, and the size of the basis set has a less pronounced effect on the total MP2 interaction energetics. This was found to be the case, as shown in Figure 3, where we plot the MP2 curves with all three different basis sets we used. It should be noted that the BSSE correction narrowed the spread of the different MP2 curves by a factor of three, pointing to substantial convergence with respect to basis set size.

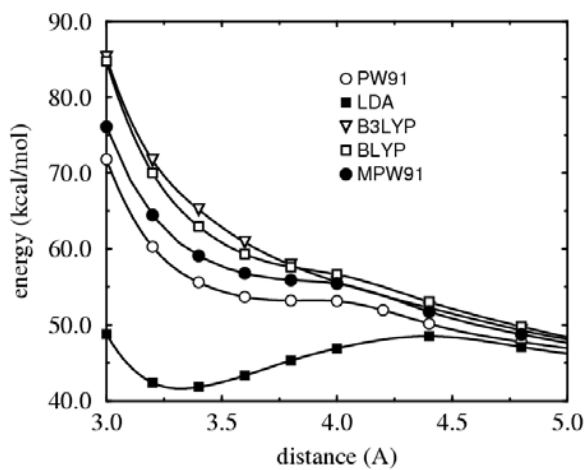

**Figure 2.** Interaction energies as a function of separation between two singly-charged bithiophene cations obtained with four DFT functionals and the 6-311+G(d,p) basis set.

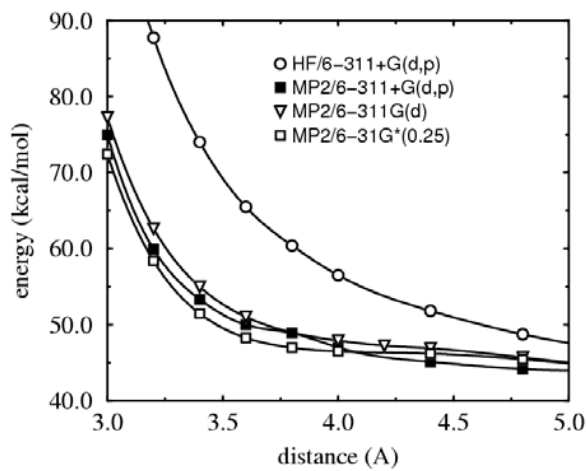

**Figure 3.** MP2 interaction energies between two singly-charged bithiophene cations obtained with three different basis sets.

To assess the convergence of the perturbative MP series, a number of MP4 calculations were performed on a test system consisting of a doubly-charged monothiophene dimer. The results were obtained at selected geometries with the 6-31G*(0.25) basis and are shown in Table 1. As expected, the interaction energies turn out to be higher than in bithiophenes (in about 15 kcal/mol), since the $\pi$-$\pi$ stabilization is weaker and the Coulombic repulsion localized in a smaller region. More importantly, the calculations exhibit only minor differences between MP2 and MP4 values.

**Table 1.** MP2 and MP4 interaction energies for two singly-charged monothiophene cations obtained at different separations with the 6-31G*(0.25) basis.

| Distance (Å) | MP2 energy (kcal/mol) | MP4 energy (kcal/mol) |
|---|---|---|
| 3.2 | 72.47 | 75.33 |
| 3.4 | 68.47 | 70.76 |
| 3.6 | 65.73 | 67.57 |

**Table 2.** Occupation numbers of the natural orbitals in the CASSCF calculations.

| distance (Å) | HOMO-3 | HOMO-2 | HOMO-1 | **HOMO** | **LUMO** | LUMO+1 | LUMO+2 | LUMO+3 |
|---|---|---|---|---|---|---|---|---|
| 3.0 | 1.91 | 1.89 | 1.78 | 1.22 | 0.78 | 0.22 | 0.11 | 0.09 |
| 3.4 | 1.93 | 1.87 | 1.82 | 1.10 | 0.90 | 0.18 | 0.13 | 0.07 |
| 3.8 | 1.93 | 1.86 | 1.84 | 1.04 | 0.96 | 0.16 | 0.14 | 0.07 |

Figure 4 depicts the orbitals involved in the CASSCF(6,4) calculations, which were picked in view of their π symmetry. Inspection of the occupation numbers reveals a significant departure from 0 or 2 only on the HOMO and the LUMO (see Table 2). The frontier orbitals are also the ones that show some changes in the occupations when the separation between monomers is increased. All other occupation numbers remain close to 1.9-2.0 or 0.1-0.0, with negligible variation as a function of distance. Such a small variation supports our use of a single-determinant approach to obtain the energy binding curves.

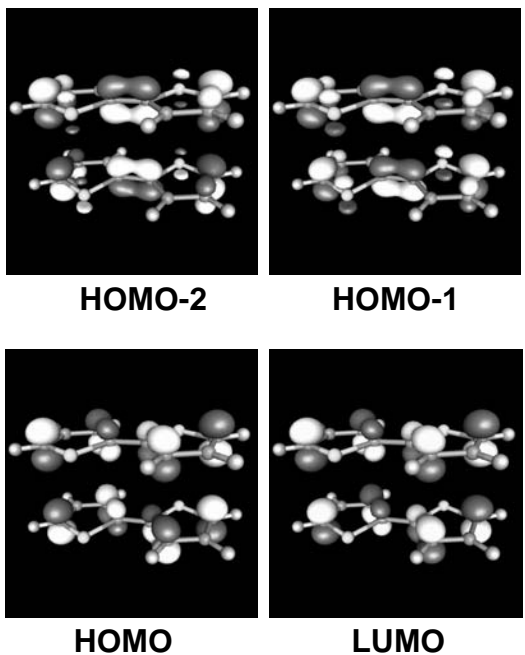

**HOMO-2**  **HOMO-1**

**HOMO**  **LUMO**

**Figure 4.** Molecular orbitals considered in the CASSCF(6,4) calculations.

None of the highly-correlated calculations supports the existence of a stable or metastable bound state. Even the remnant shoulder present in PW91 disappears in some of the other DFT profiles we explored (BLYP, B3LYP), which yield a repulsive behavior in all the distance range. The role of electron correlation is nevertheless readily apparent, contributing an attractive term to the interaction energy that is especially relevant for separations below 4 Å. At larger distances the π-π overlap virtually disappears and the interactions become mainly electrostatic, to the point that beyond 5 Å all curves agree. We considered also the possibility of a bipolaronic stabilization, where two holes bind to one distorted oligomer and the other remains uncharged, but it was ruled out by the calculations, since such asymmetric charge separation turns out to be unstable in comparison to a pair of singly-charged monomers.

In conclusion, we found the interaction energies in this system are repulsive for all distances, establishing that doubly-charged dimers of bithiophenes are neither stable nor metastable in vacuum. While LDA-DFT provides a very poor and qualitatively inaccurate picture of the pairing interactions, more accurate DFT approaches and quantum-chemistry methods are in qualitative and largely quantitative agreement. An accurate treatment of electronic correlations is needed to get close to thermodynamic accuracy; nevertheless all methods presented, with the exclusion of DFT-LDA, describe a consistent picture of repulsion between the dimers. Given the similarities in the DFT calculations for bi-, ter-, and quaterthiophenes, we believe this conclusion to be of general applicability for all of the short oligothiophenes. Intrinsic factors such as π-π bonding and electron delocalization, always present in the gas phase and conventionally considered as the driving forces for the stacking, are not sufficient to explain the dimerization observed in the condensed phase. While for an isolated dimer of oligothiophenes the electrostatic repulsion will always prevail on the attractive π-π interactions, in the solid state the effects of counterions, packing constraints, or even mixed valence states may well determine the stabilization of the stacked form.

In fact, much of the experimental evidence comes from data in solution. To study the role of solvation, we have explored the effects that the presence of a self-consistent polarizable medium would have on the delicate balance between Coulombic repulsion and covalent interactions. For our calculations we focused on dimers of bi- and ter- thiophenes, using GGA-PW91 and embedding the quantum fragments in a continuous polarizable solvent employing the IPCM method.[16] The dielectric constant has been set to that of acetonitrile, since experimental data is most often reported in this medium. As it is readily apparent from Fig. 5, solvation does dramatically affect the energetics of dimerization, leading now to an effective attractive interaction at short distances, and to the appearance of a bound state, thermodynamically stable with respect to separation. This result can be qualitatively understood recalling that a polarizable solvent favors the concentration of charge in smaller cavities. This effect enhances the stability of bound states with respect to the dissociated constituents, and in the present cases reverses the balance from overall repulsion to attraction and stability. The curves in Fig. 5 show the binding energy for dimers of singly-charged bi- and terthiophenes in GGA-PW91 and acetonitrile, obtained as the energy of the dimer minus two times the energy of the monomer. For the case of terthiophenes we find a thermodynamically stable state with a binding energy of 5.2 kcal/mol; the equilibrium separation of 3.4 Å is very close to that observed in π-stacks in the solid state,[1d] and represents a remarkably large effective interaction distance. Dimerization is also found in singly-charged bithiophenes, with a slight decrease in the binding energy of 1.5 kcal/mol, and in agreement with experimental evidence showing a direct relation between the length of oligomers and the tendency to dimerize.[1c] Such relation can be ascribed to the increase in $\pi$-$\pi$ overlap accompanied by a 'dilution' of the Coulombic repulsion as the size of the oligomer increases. The role of non-electrostatic solvation terms (cavitation, dispersion and repulsion) on the overall stabilization is negligible, since in the dimer they are roughly twice as large as in the monomer. Experimental values for the exact structures explored in these simulations are not available, but studies for bi- and terthiophene derivatives report interaction enthalpies in the range of 4 to 15 kcal/mol in dichloromethane and acetonitrile.[1a,b, 20]

Figure 5 also shows the preferred geometries for the dimers in solution: bithiophenes align in an eclipsed configuration, while terthiophenes switch to the same slipped arrangement found for quaterthiophenes in vacuum. A configuration search at a fixed separation distance was performed to determine the arrangement that minimizes the interaction energy for the solvated dimers. For these calculations the 6-31G(d) basis was used; we show in Figure 6 the variation of the energy as one monomer is slipped over the other along the molecular axis. The interaction energy is very sensitive to these displacements, and can be positive for a broad range of values, more so for the shorter oligomer. The optimal arrangement for the terthiophene dimer was found in correspondence of a shift of 2.05 Å on the molecular axis, and for a simultaneous displacement in the perpendicular direction of 0.2 Å. These structural shifts reveal the underlying pattern in the nodal surface, that remains almost unaffected by the solvent. The bithiophene dimer is found most stable in the eclipsed geometry. These optimal alignments were used to construct the curves in Figure 5, varying the interchain separation.

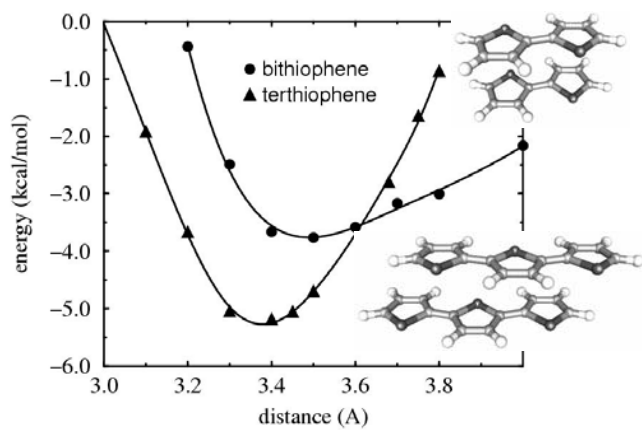

**Figure 5.** Total energy as a function of separation between two singly-charged oligothiophene cations solvated in CH$_3$CN. The minimum energy structures are displayed on the right.

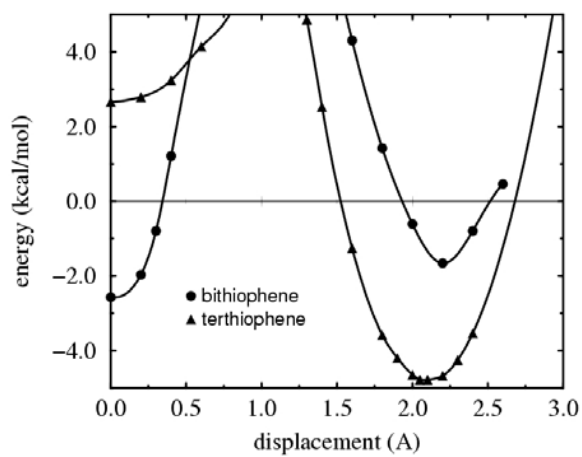

**Figure 6.** Interaction energy for the bithiophene and terthiophene dimers as calculated with PW91 and IPCM, as a function of the shift along the molecular axis. The distance between chains was kept constant at 3.4 Å.

**Conclusions**

We presented here accurate ab-initio calculations on oxidized oligothiophene dimers, both in vacuum and in solution, using several quantum-chemistry approaches. Our results do not support the existence of stable or metastable dimers for small oligomers in vacuum; π-π interactions between singly oxidized oligothiophenes are not strong enough to overcome the Coulombic repulsion. In a polarizable solvent (acetonitrile) we found stable doubly-charged π-dimers for both bi- and terthiophenes, with binding energies comparable to experimental findings. These results highlight the delicate balance between electrostatic and correlation effects in determining the structural and electronic stability of these systems.

**ACKNOWLEDGMENT**. We would like to thank T. Swager and I. Hunter for sharing their expertise on oligothiophenes, and C. Filippi, P. Anquetil, and H.-H. Yu for very valuable discussions. This research was supported by ONR grant N000014-01-1-1061 and by the Institute of Soldier Nanotechnologies, contract DAAD-19-02-D0002 with the U.S. Army Research Office.